\documentstyle[preprint,aps]{revtex}

\begin{document}
\draft
\title{\bf Selfgravitating nonlinear scalar fields}
\author{Edward Malec}
 \address{Institute of Physics, 30-59 Krak\'ow, Reymonta 4}

\maketitle
\begin{abstract}
 We investigate the Cauchy problem for the Einstein - scalar field equations
   in asymptotically flat spherically
symmetric spacetimes, in the standard $1+3$ formulation.
We prove the local existence and uniqueness
of solutions  for initial data
given on a space-like hypersurface
in the Sobolev $H_1\cap H_{1,4} $ space.
Solutions exist globally if a central (integral) singularity does  not form
and/or outside an outgoing null hypersurface.  An explicit example
demonstrates that there exists a local evolution with a naked
initial  curvature singularity at the symmetry centre.
\end{abstract}

\centerline{I. Introduction.}

The local Cauchy problem in General Relativity has been solved long ago by
Y. Choquet - Bruhat \cite{Choquet - Bruhat} in the so-called harmonic gauge
 but the global Cauchy problem still remains unsolved,
despite progress made in recent years.
The list of known results concerning evolution
 in asymptotically flat spacetimes includes the global existence of
almost Minkowskian geometries \cite{Chr-Kl}
and two special cases of spherically symmetric  systems - massless
scalar fields with  characteristic initial data  \cite{ (Christodoulou) }
and the Vlasov - Einstein equations \cite{rendall95}.
On the other hand the validity of the main open question of gravitational
physics, the cosmic censorship  hypothesis \cite{Penrose}, would demand
the existence of global Cauchy  solutions.  More radically, the cosmic
censorship question  can be identified with  the global Cauchy
problem\cite{Moncrief}.

In this paper we consider the Cauchy problem  for Einstein equations
coupled to a class of nonlinear scalar fields.
  We specialize to spherically
symmetric and asymptotically flat systems, with initial data prescribed on
a space-like hypersurface.  Our  interest is in finding the
weakest possible solutions; that  is motivated by the existence of an
$L_2$ apriori estimate induced (in the absence
of black holes) by the conservation of
the asymptotic mass in   asymptotically
flat spacetimes.  An ultimate  reduction to the $L_2$ class would mean
that the global evolution exists in the absence of black holes.
 We   did not  achieve that aim  although the differentiability
  of solutions considered here is $H_1\cap H_{1,4} $, i. e., weaker
than of classical $C^1$ solutions.  We show elsewhere \cite{Malec1996}
that the breakdown of the evolution, i. e., 
the lapse collapse at the
symmetry center,  must be associated with  the infinite value of the
$H_1$ norm of a solution and with the conical singularity.

The plan of this paper is the following. Section  II
presents the Einstein - scalar field equations. They can be reduced
to a system of two first order characteristic equations on $[0, \infty )
\times [0, T]$. Further analysis of   metric coefficients of the equations
allows one to reduce that to a single "symmetrized" equation on
$(-\infty , \infty )\times [0, T]$.
Section III comprises a number of
estimates that will be used in further sections. Section IV consists
of the main local result, Theorem 6.  The local existence is proved
in a standard way, by a combination of the "viscosity" and compactness
methods.  The global uniqueness is  shown in Section V.
Section VI presents a  proof of a related global  Stefan problem, i. e., that
a global Cauchy evolution exists outside an event horizon and, in particular,
the Schwarzschild radius $R=2m$.
Section VII shows the global existence, assuming that a "central
integral singularity" does not exist.    Section VIII presents an example of
initial configuration for the  scalar field
with a    singularity at the symmetry center that can be  seen by external
observers, i. e., it is naked. That demonstrates, in our opinion,  that  the
concept of {\it pointwise  singularities} shall be replaced
  by a smaller class of {\it quasilocal (integral) singularities}
and that the concept of the cosmic censorship shall be accordingly
reformulated.

  \vskip 2cm

\centerline{II. Equations.}

In spherically symmetric spacetimes one can always choose a diagonal
line element
 \begin{equation}
ds^2 = -  N^2(r, t)dt^2 + a(r, t) dr^2 +  R^2(r, t)  d\Omega^2,
\label{2.1}
\end{equation}
where  $t$  is a time coordinate, $r$  - a radial coordinate, $R $
 - an  areal radius and $d\Omega^2 = d\theta^2 + \sin^2\theta d\phi^2$ -
 the   line element on the unit sphere, $0\le \phi < 2\pi $
and $0\le \theta \le \pi $.
 At spatial infinity $N=1$ and $a=1$, for asymptotically flat spacetimes.
Below we adopt the standard convention that Greek letters change from 0 to 3
while Latin indices range from 1 to 3.

Einstein  equations $R_{\mu \nu }-{g_{\mu \nu }\over 2}R=8\pi T_{\mu \nu }$
can be written    as a $1+3$ system of initial constraints
and evolution equations \cite{Wald}.
Let       $\Sigma_t $ be a foliation of Cauchy hypersurfaces, with
   $g_{ij}$ the intrinsic metric  and $K_{ij}$ the extrinsic curvature.
We adopt the convention of Wald \cite{Wald} so that the metric signature
is $(-+++)$,
  $K_{ij}={\partial_tg_{ij}\over 2N }$ and $trK =g^{ij}K_{ij}$.
  Let  $T{\mu \nu }$  be  the energy - momentum tensor of a matter field.
The matter energy density is $\rho =-T_0^0 $  and  the matter current
density  reads  $j_i=NT_i^0$.

 Then the Einstein  constraints read:
 \begin{equation}
^{(3)}R-K_{ij}K^{ij}+(trK)^2=16\pi \rho
 \label{2.2}
\end{equation}
 \begin{equation}
 \nabla_iK^{ij}-\nabla^jtrK =-8\pi j^j
 \label{2.3}
\end{equation}
Above   $^{(3)}R$  is the scalar curvature of the intrinsic metric of
  $\Sigma_t $. It is useful to express the Einstein equations in terms
of the mean curvature of  a two-dimensional   sphere centered around
the symmetry center  of $\Sigma_t $, $p={2\partial_rR\over  \sqrt{a} R}$,
and the following components of the extrinsic curvature:
 \begin{equation}
trK-K =2 K_{\theta }^{\theta } =2 K_{\phi }^{\phi },~~~~K=K_r^r.
\label{2.6}
\end{equation}

The   constraints in terms of $K$ and $p$  read
\begin{eqnarray}
{\partial_r (  p R )\over \sqrt{a}}= -8\pi R\rho  -{3R\over 4}(K)^2
+{R\over 4}(trK)^2+{R\over 2}KtrK -{Rp^2\over 4} +{1\over R}
\label{2.7}
\end{eqnarray}
\begin{eqnarray}
{\partial_r ( R^3( K -trK) )\over \sqrt{a}}= -8\pi R^3 {j_r\over \sqrt{a}}
-ptrKR^3
\label{2.8}
\end{eqnarray}

 The two remaining equations are the evolution Einstein equation:
 \begin{eqnarray}
 \partial_0(  K -trK )= {3 N \over 2}(  K)^2 +{N\over 2}(trK)^2-2NKtrK
 -{ p^2R\over \sqrt{a}}\partial_r { N \over pR}+8\pi N(T_r^r +\rho )
 \label{2.9}
 \end{eqnarray}
and the lapse equation:
 \begin{eqnarray}
\nabla_i \partial^i  N  = N \Bigl( {3  \over 2}(  K)^2  + {(trK)^2\over 2}
-KtrK +4\pi  (\rho + T_i^i)    \Bigr) +\partial_0trK .
 \label{2.10}
\end{eqnarray}
The above equations yield (via the Bianchi identity)
the energy-momentum conservation equations:
 \begin{eqnarray}
 \partial_0{j_r\over \sqrt{a}} +N(trK +K) {j_r\over \sqrt{a}}+
 {N\over \sqrt{a}}\partial_r T_r^r
 +{\partial_rN\over \sqrt{a}}(T_r^r+\rho)+Np(T_r^r-T_{\phi }^{\phi })=0
 \label{2.11}
\end{eqnarray}
 \begin{eqnarray}
- \partial_0\rho -{Np\over \sqrt{a}} j_r-{N\over \sqrt{a}}\partial_r
({j_r\over
\sqrt{a}}) -{2\partial_rN\over a}j_r-NK(T_r^r-T_{\phi }^{\phi })
-NtrK(\rho +T_{\phi }^{\phi })=0.
 \label{2.12}
\end{eqnarray}
The above equations can be converted to a system of nonlinear integral
equations. They are particularly simple in the so-called polar gauge
     $trK=K$.    By solving the hamiltonian constraint one obtains
\begin{equation}
pR=2\sqrt{1-{2m\over R}+{2m(R)\over R}},
\label{3.1}
\end{equation}
where $m$ is the asymptotic mass and $m(R)=4\pi \int_R^{\infty }drr^2\rho $
and from the evolution equation
\begin{equation}
N={pR\over 2} \beta (R),
\label{3.3}
\end{equation}
\begin{equation}
\beta (r) = e^{16\pi \int_r^{\infty }(-T_r^r-\rho )
{1\over p^2s}ds}.
\label{3.4}
\end{equation}
 The line element reads
\begin{equation}
ds^2= -dt^2N^2  +{4\over (pR)^2}dR^2 +R^2d\Omega^2.
\label{3.5}
\end{equation}
The above equations mean   that   metric functions can be expressed
as some integrals of matter fields and that spherically symmetric
Einstein equations do not exhibit any dynamical meaning. The
whole dynamics of a selfgravitating spherical system is contained
in the evolution of  the material field.

The nonlinear scalar field equation is given by  the second order equation
\begin{equation}
\nabla_{\mu }\nabla^{\mu }\phi - W'(\phi )=0,
\label{4.1}
\end{equation}
where $W(\phi ) $ is the scalar field potential and $W'(x)= {d\over dx }W(x)$.
They can be written in the characteristic form, as a system of two
first-order equations,
{\bf \begin{eqnarray}
(\partial_0 +{NpR\over 2}\partial_R)V  = {8\pi N\over p}V (j-T) -
{Np\over 4} V-{NUp\over 2}  -{NV \over pR^2} + N W',
\label{6.5}
\end{eqnarray}
and
 \begin{eqnarray}
(\partial_0 -{NpR\over 2}\partial_R)U  = {8\pi N\over p}U (j+T) +
{Np\over 4} U+{NVp\over 2}   +{NU \over pR^2} - NW';
\label{6.6}
\end{eqnarray}   }
above
\begin{eqnarray}
&&V =\underline D \phi \nonumber\\
&&U =D  \phi \nonumber\\
&&j=N{T_r^0\over \sqrt{a}}={1\over 4}(V^2-U^2)\nonumber\\
  &&T=T_r^r=  {1\over  4}(V^2+U^2)- W(\phi ). \nonumber\\
 &&\rho  = {1\over  4}(V^2+U^2)+ W(\phi ),
\label{6.7}
\end{eqnarray}
where
\begin{eqnarray}
&&\underline D={-1\over N}\partial_0 +{pR\over 2}\partial_R\nonumber\\
&&  D={1\over N}\partial_0 +{pR\over 2}\partial_R
\label{6.7a}
\end{eqnarray}
 Equations (\ref{6.6}) and (\ref{6.7}) are
 hyperbolic   in a " strict sense"  if $NpR>0$  i. e. $p >0$.

Define
\begin{eqnarray}
&&\phi ={\tilde \phi  \over R} \nonumber\\
&&h_+ ={\underline D \tilde \phi \over pR} \nonumber\\
&&h_- ={ D \tilde \phi \over pR}\nonumber\\
&&\hat h= {1\over 2R}\int_0^Rdr(h_+ +h_-)
\label{6.8}
\end{eqnarray}
Notice that $\tilde \phi  =  \int_0^Rdr(h_+ +h_-) $, since by
continuity one has to impose $\tilde \phi (0)=0$.

Define also
\begin{equation}
\delta (R)  = {NpR\over 2}
\label{6.9}
\end{equation}
Differentiation of (\ref{6.9}) gives, with the help of the hamiltonian
constraint (\ref{2.7}), the relation $\partial_R\delta =
{\beta -\delta \over R} -8\pi WR\beta $. That allows one  to express $\delta $
in the following useful form
\begin{eqnarray}
\delta (R)={1\over R}\int_0^R\beta dr-{8\pi \over R}\int_0^Rdrr^2\beta
W(\hat h).
\label{6.10}
\end{eqnarray}
One can show, after some calculations,  that
\begin{eqnarray}
 \beta (R) = e^{-8\pi \int_R^{\infty }{dr\over r} \bigl(
(h_+ -\hat h)^2 +(h_- -\hat h)^2\bigr) }
 \label{6.10a}
\end{eqnarray}
{\bf Remark. } One can easily show, by analyzing the
Einstein constraints,
that if a collapsing system possesess
apparent horizons, then at least one of them (the innermost  apparent horizon)
  must be immersed inside matter. In the polar gauge apparent horizons
  coincide with minimal surfaces, i. e., surfaces at which the trace $p$ of
the second fundamental form  vanishes.
Formulae (\ref{6.10} and \ref{6.10a}) imply now that derivatives of
  metric functions $\beta , \gamma  $ and $\delta $
must be singular at the location of the innermost minimal surface.
That    means  that smooth solutions  can exist only if minimal surfaces
are absent, that is, the system of equations is strictly hyperbolic.

The scalar field equations reduce now to two  first order  equations
\begin{eqnarray}
&&(\partial_0 +\delta \partial_R)h_+  = (h_+ -\hat h )
(8\pi \beta RW+ {\gamma \over R} )  +{\beta R\over 2}W' \nonumber\\
&&(\partial_0 -\delta \partial_R)h_-  = -(h_- -\hat h )
(8\pi \beta RW+ {\gamma \over R} )  -{\beta R\over 2}W',
\label{6.11}
\end{eqnarray}
where $\gamma (R)= \delta (R)- \beta (R)$.
Let us define a function by $h(R)=h_+(R) $ for $R>0$ and
$h(R)=h_-(-R) $ for $R\le 0$. Then one can write down functions
$\hat h$ and  $\beta $   as follows
\begin{eqnarray}                   
 &&\hat h= {1\over 2R}\int_{-R}^Rdrh(r)  \nonumber\\
 &&\beta (R) = e^{-8\pi (\int_R^{\infty }+\int_{-R}^{-\infty })
 {dr\over r}   (  h -\hat h)^2    }.
 \label{6.12}
\end{eqnarray}
>From  that follows  $\hat h(R)=\hat h(-R)$,  $\beta (R)=\beta (-R)$ and
      $W(R)=W(-R)$.
>From   (\ref{6.10})  one infers  that  $\delta (R) = \delta (-R)$
 and that implies  $\gamma (R) =\gamma (-R)$.
Therefore the system of two
first order  equations  (\ref{6.11}) can be written as a single first
order equation on a "symmetrized" domain $-\infty \le R\le \infty $,
\begin{eqnarray}
&&(\partial_0 +\delta \partial_R)h  = (h -\hat h )
(8\pi \beta RW+ {\gamma \over R} )  +{\beta R\over 2}W'.
\label{6.13}
\end{eqnarray}
That is the central equation of this paper; together with definitions
of $h, \hat h , \beta , \delta $ and $\gamma $ it   encodes the all
information carried by Einstein equations coupled to the scalar field.
Notice that $\int_{-\infty }^{\infty }dr h(r)=\int_0^{\infty }dr( h(r)+h(-r))=
\int_ 0^{\infty }dr \partial_r\tilde \phi =0$; therefore initial data
$h_0$ of compact support  must satisfy the condition
\begin{equation}
 \int_ {-\infty }^{\infty }dr    h_0 =0.
\label{h_0}
\end{equation}
 One can easily show, using
relations between metric functions and their symmetry properties, that if
(\ref{h_0}) holds true then $\int_{-\infty }^{\infty }dr h(r, t)=0$ in the
existence interval of a solution.

In Theorem 6 of Section 4 we formulate and then prove the  local existence
of solutions of (\ref{6.13}). Its uniqueness is shown in Theorem 8.
Let us point out that the above equation incorporates some of sigma
models (but let us point out that
  the local existence result of Theorem 6
does not apply to them).
The description of  selfgravitating Yang - Mills $SU(2)$  potentials
reduces also to a single equation of a similar form.
\vskip 2cm

\centerline{III. Estimates of metric functions and of $\hat h(R)$.}

We define   Sobolev space $H_1(V)$ - as a completion
of $C^1$-functions in the norm $||f||_{H_1(-\infty , \infty )}=
\Bigl( \int_{-\infty }^{\infty }dR (\partial_Rf)^2\Bigr)^{1/2}$.
This Section contains a
number of estimates that will be used later in order to prove the
local existence and uniqueness of solutions.

 Define
 \begin{equation}
\hat h(R) = {1\over 2R}\int_{-R}^Rh(r)  dr.
\label{l1}
\end{equation}
{\bf Lemma 1.} Let $h(R)\epsilon H_1(-\infty ,\infty )$, $ h(r)  $
be of compact support  with $h(r)=0$ for any $|r|>R_0$ and let $1> \delta >0$.
Then $|\hat h|\le C||h||_{H_1(-\infty , \infty )}$,
$\hat h(R) \epsilon L_2(0 ,\infty )$  and
$$||r^{\delta }\partial_r\hat h||_{L_2(-\infty  ,\infty )}\le {2R^{\delta }_0
\over \delta } ||h||_{H_1(-\infty , \infty )}.$$
Proof.

{\bf Step 1}.

 $|h(R)-h(0)|\le  ||h||_{H_1(-\infty , \infty )} {R^{1/2}\over \sqrt{2}}$,

 $|h(0)|\le ||h||_{H_1(-\infty , \infty )} {R^{1/2}_0\over \sqrt{2}}$.

Proof of Step 1.

$h(r)$ is of class $H_1(-\infty  ,\infty )$, that is $h \epsilon C^{1/2}$.
Now we have
$$| h(R)-h(0)|=|\int_{-R}^Rdr\partial_rh(r) |\le
||\partial_r h||_{L_2(-\infty ,\infty )}{R^{1/2}\over \sqrt{2}}\le  $$
$$||h||_{H_1(-\infty , \infty )} R^{1/2},$$
where the first inequality follows from the Schwartz inequality.
For a function $h$ vanishing outside a region $|R|\le R_0$
one obtains $h(0) \le ||h||_{H_1(-\infty , \infty )}
{R^{1/2}_0\over \sqrt{2}}$.
That proves   Step 1.

{\bf Step 2}.

 $|\hat h(R)- h(0)| \le R^{1/2}  ||h||_{H_1(-\infty , \infty )}$;

Proof of Step 2.

 Notice the identity
$$|\hat h(R)-h(0)| = |{1\over 2R} \int_{-R} ^R(h(r)-h(0))dr |$$
and   use  the estimation   of Step 1. That immediately yields
$|\hat h(R)-h(0)|\le {R^{1/2}\over 3\sqrt{2}}
||h||_{H_1(-\infty , \infty )} $.

{\bf Step 3}.

$$|\partial_R\hat h(R) | \le 2R^{-1/2} ||h||_{H_1(-\infty , \infty )}.$$

Proof of Step 3. From (\ref{l1}) one gets
$$\partial_R\hat h(R)= {-1\over 2R^2}\int_{-R}^Rdr(h(r)-h(0))
+{1\over 2R}(h(R)+h(-R)-2h(0)).$$
Using Step 1 and performing simple integrations, one arrives at
$$ |\partial_R\hat h(R) |\le {2\over \sqrt{R}}
||h||_{H_1(-\infty , \infty )}.$$
Proof of Lemma 1.
 Estimations follow directly  from  definitions of corresponding
norms and from  Steps 1 - 3.  One has to use   the assumption that
 a support of $h(R)$ is finite, which gives $\hat h(R)={C\over R}$ outside
the support of $h$; that ensures the $L_2$ integrability of $\hat h$.

Define
\begin{equation}
<h>= h-\hat h.
\label{h1}
\end{equation}
{\bf  Lemma 2.} Let $h$ satisfies conditions of Lemma 1 and
$1>\eta >0$. Then
\begin{eqnarray}
&&|<h>|\le CR^{1/2}||h||_{H_1(-\infty , \infty )}\nonumber\\
&&||R^{\eta }\partial_R<h>||_{L_2(-\infty , \infty }
\le C||h||_{H_1(-\infty , \infty )}
\label{h2}
\end{eqnarray}
Proof of Lemma 2.

Notice that $<h>(R) = h(R)-h(0) -{1\over 2R}\int_{-R}^Rdr (h(r)-h(0))$
where $|h(r)-h(0)| $ is bounded by   Step 1. That gives the
first estimate of Lemma 2.

 The second, integral, bound on
$$\partial_R<h>= \partial_Rh - \partial_R\hat h$$
follows immediately from  Lemma 1.

Define
$$\beta (R)=e^{-8\pi (\int_R^{\infty }+\int_{-R}^{-\infty })dr
{1\over r}   <h>^2 }$$
{\bf  Lemma 3.} Let $h$ satisfies conditions of Lemma 1.
 Then

i) $e^{- C||h||_{H_1(-\infty , \infty )}^2}\le \beta (R) \le 1,$

ii) $|\partial_R\beta (R)| \le C ||h||_{H_1(-\infty , \infty }^2$  and
$\partial_R\beta (R) |_{R=0}=0$,

with $C$'s being some constants depending  on the support of $h(R)$.

Proof.

i) Obviously $\beta (R) \le 1$.
The lower bound of i) follows from the first estimate of Lemma 2 on
$<h>$. Invoking to the   finiteness of
the support  of $h(r)$ and $\hat h(r)$, one arrives at the sought inequality.

ii) Direct differentiation of $\beta $ with respect $R$ yields
$${d\over dR}\beta (R) ={8\pi  \over R}\Bigl(  <h>^2(R)
+ <h>^2(-R) \Bigr) \beta (R).$$
The first estimate of  Lemma 2 yields
$|\partial_R\beta (R)| \le C ||h||_{ H_1(-\infty , \infty )}^2$.
>From   Lemma 2 we have $|< h (R)>|\le R^{1/2}||h||_{ H_1(-\infty ,
\infty )}$; $\partial_R\beta (R) $ is continuous for $R  \ne 0$
and, being an antisymmetric function of $R$, must vanish at the origin.
That gives   ii).

Define
\begin{equation}
\gamma (R)={1\over R}\int_0^R\beta dr-\beta (R)
-{8\pi \over R}\int_0^Rdrr^2\beta W(\hat h)
\label{6.gamma}
\end{equation}
{\bf Lemma 4.} Let $h$ satisfies conditions of Lemma 1. Assume
that $|W(x)|$ can be bounded from above by  a polynomial of k-th order in $x$
with constant coefficients, and $W(x)\ge 0$. Then

i) $ | \gamma(R)| \le CR
(||h||_{H_1(-\infty , \infty }^2+
||h||_{ H_1(-\infty , \infty }^{k}),$

ii) $|\partial_R\gamma (R)| \le C (||h||_{ H_1(-\infty , \infty }^2+
||h||_{H_1(-\infty , \infty }^{2k}),$

where $C$ changes from a line to line, but it depends only on $R_0$  and
coefficients of $W$. Morever, $\partial_R\gamma (0)=0$.

Proof of Lemma 4  is straightforward and consists in applying hitherto
proven estimates  in order to bound the derivatives of $\gamma $ in question.

i) Notice that
\begin{eqnarray}
&&|\gamma (R)| = |{1\over R}\int_0^R(\beta (r) -
\beta (0))dr -\beta (r)+\beta (0)
-{8\pi \over R}\int_0^Rdrr^2\beta W(\hat h)|= \nonumber\\
&& |{1\over R}\int_0^Rdr \int_0^rds \partial_s\beta (s) -
\int_0^Rds \partial_s\beta (s)-{8\pi \over R}\int_0^Rdrr^2\beta W(\hat h)|\le
\nonumber\\
&&CR(||h||_{H_1(-\infty , \infty }^2+
||h||_{ H_1(-\infty , \infty }^{k}),
\end{eqnarray}
where in the last line we used the estimation ii) of Lemma 3.

Using the mean value theorem, one can
write the second line of the preceding equation   as
\begin{equation}
-\int_{\theta R}^R\partial_r\beta dr -
{8\pi \over R}\int_0^Rdrr^2\beta W(\hat h ),
\end{equation}
where $1>\theta >0$. From that and from the estimation ii) of  Lemma
3 one arrives at the second estimate of Lemma 4.  By antisymmetry and
continuity of $\partial_R\beta $
we have also $\partial_R\gamma (0)=0$.  That accomplishes the proof of
Lemma 4.

Define
$$\delta (R) = \gamma (R) +\beta (R);$$
estimates of derivatives of $\delta $ up to  first order
follow immediately from those of $\gamma $ and $\beta $.  Thus

{\bf Lemma 5.} Let $h$ satisfies conditions of Lemma 1. Assume
that $|W(x)|$ is bounded by a polynomial of k-th order in $x$,
 and $W(x)\ge 0$. Then

i) $  \delta(R) \le CR (||h||_{H_1(-\infty , \infty }^2+
||h||_{H_1(-\infty , \infty }^{k}),$

ii) $|\partial_R\delta (R)| \le C (||h||_{H_1(-\infty , \infty }^2+
||h||_{H_1(-\infty , \infty }^{2k}),$

and $\partial_R\delta |_{R=0}=0$.
 Above $C$ changes from a line to line, but it depends only on $R_0$  and
coefficients of $W$.

\vskip 2cm

\centerline{IV. The existence of local Cauchy solutions.}

{\bf Definition.} We define $H_{1,4}(a, b)$ as a completion of classical $C^1$
functions $f$ of compact support in the $||\partial_rf||_{L_4(a, b)}$ norm.

In the case of $a,b <\infty $  we have the inclusion $H_{1, 4}(a, b)
\subset H_1 (a, b)$.

We will frequently use the following technical  result.

{\bf Proposition A.}
Let $ f$ be a  continuous function of a compact support
$\Omega = [-R_0, R_0]\times (0, T)$ and A, B some constants (depending
on $R_0$ and $T$) such that for all $0\le t\le T$

i)   $||f||_{H_1 ([-R_0, R_0]  )}<A$.

ii)  $||\partial_0f||_{L_2([-R_0, R_0] )}<B$.

Then  there exists a constant $C$ depending only on $R_0$ and $T$
such that
$$|f(R_1, t_1)-f(R_2, t_2)|\le C(|R_1-R_2|^{1/2}+|t_1-t_2|^{1/2}).$$
 For the proof see \cite{Godunov}. Below we shall outline its main points.
 A part of the above statement, the "equal time  
 inequaity", can be proven in a way similar to that employed
 in  Step 1. Similarly as before 
 one shows that $|f(R_1, t)-f(R_2, t)|\le ||f||_{H_1}\sqrt{|R_1-R_2|}$. 
 The compactness of the support of $f$ yields then $\sup |f|
 \le  ||f||_{H_1}\sqrt{|2R_0|}\le A\sqrt{|2R_0|}$. Now notice that for any
 $-R_0\le R_1 \le R_2 \le R_0$ and $0<t_1<t_2<T$
 \begin{eqnarray}
 &&\int_{R_1}^{R^2}dr\Bigl( |f(r, t_1)- f(r, t_2)|\le \int_{t_1}^{t_2}
\int_{R_1}^{R^2}dr|\partial_tf|\le \nonumber\\
&&B(t_2-t_1)\sqrt{(R_2-R_1)};
\label{G1}
\end{eqnarray}
we use the Schwartz inequality and the assumption ii).

By continuity of $f$ there exists a point $R_3$ lying between $R_1$ and $R_2$
such that $|f(R_3, t_1)- f(R_3, t_2)| \le B(t_2-t_1)\sqrt{{1\over 
(R_2-R_1)}}.$ Notice that $|f(R_1, t_1)-f(R_1, t_2)|\le
|f(R_1, t_1)-f(R_3, t_1)|+ |f(R_3, t_2)-f(R_1, t_2)|+
|f(R_3, t_1)-f(R_3, t_2)|$;
 employing the "equal time"  inequality for
the function $f$ at fixed times $t_1$ and $t_2$ and choosing 
${t_2-t_1\over T} = {R_2-R_1\over 2R_0}$ one arrives at 
$|f(R, t_1) -f(R, t_2)| \le C\sqrt{t_2-t_1}$ for any $R\epsilon {R_1, R_2}$. 
Combining that result with the "equal time" inequality one accomplishes the
proof of Proposition A.

 {\bf Theorem 6.} Let the initial data of  equation (\ref{6.13})
 on an initial slice $\Sigma_0$ be  of compact support,
 $\inf_{\Sigma_0} \delta >0$ and

i) $  h_0\epsilon H_{1,4}(-\infty , \infty )$; assume also that

ii) $\int_{-\infty } ^{\infty } dr h_0(R, t)=0$.
 Let $0\le W(x)$  and $|W'(x)|$ be bounded by a polynomial with constant
 coefficients of order $k $. Then there exists a local Cauchy  solution
 of (\ref{6.13}).

 \vskip 0.5cm
 {\bf Remark.} Theorem 6 implies the existence   of
a foliation $\Sigma_{t} $ for some $ T $( $0\le t < T$), with $ h_+,
  h_- \epsilon H_{1}(0, \infty )$,  and with no minimal
 surfaces on any leaf $\Sigma_{t} $.  Indeed, having $h(R, t)$, one determines
all metric functions and the scalar field itself - see Section II for
corresponding formulae.  That minimal surfaces are absent in a local
evolution follows from the proof, where the positivity
of $ \delta $ is proven. The  existence of a local evolution of (\ref{6.13})
 can be proven without the assumption ii); the latter is needed to make  the
 identification with the Einstein - scalar field equations (see  a remark
 at the end of Section 2).

 {\bf Proof.}

Let us notice that   Eq.  (\ref{6.13})  is   nonlocal and
integro-differential.  We prove its solvability from    first principles.

In the part A of the proof we consider  a regularized equation
  in $H_1$. The existence of a local in time solution
  is proven in a standard way,  using a method of succesive
appoximations and then   standard compactness method.

In the part B we show that if initial data are  in $H_{1,4}$, then the
regularization can be removed. Once again the compactness method
  ensures  the  existence of a weakly
 convergent  subsequence,  whose limit is  the sought (local in time)
 solution of the reduced equations (\ref{6.13}).
\vskip 0.5cm
{\bf Part A.}

 Let us define  a regularized equation,
\begin{eqnarray}
 (\partial_0 +\delta \partial_R)h   =
<h> \Bigl( 8\pi \beta RW+ {\gamma |R|^{\epsilon }\over R} \Bigr)
+{\beta R\over 2}W',
\label{ 7.1}
\end{eqnarray}
where all coefficients are defined as in section II.   (The introduction
of the parameter $\epsilon $ is reminiscent of the viscosity method
known in the Navier - Stokes equation.)
Denote  a solution  of (\ref{ 7.1}) by $h_{\epsilon }$.
Define a sequence of functions $h_{n\epsilon }(t, R)$ as follows:
$$h_{0\epsilon }(t, R)=h(t=0, R)$$
and $h_{n\epsilon }$ is a solution of
\begin{eqnarray}
&&(\partial_0 +\delta_{n-1} \partial_R)h_{n\epsilon }  =
<h_{n-1}>
(8\pi \beta_{n-1} RW_{n-1}+ {\gamma_{n-1}|R|^{\epsilon } \over R} )  +
{\beta_{n-1}R\over 2}W'_{n-1}, \nonumber\\
\label{7.1a}
\end{eqnarray}
where
\begin{eqnarray}
&&\beta_n (R) = e^{-8\pi (\int_R^{\infty }+\int_{-R}^{-\infty })
{dr\over r}    < h_n >^2}   \nonumber\\
&&\delta_n (R)={1\over R}\int_0^R\beta_n dr-{8\pi \over R}
\int_0^Rdrr^2\beta_n W_n(\hat h_{n\epsilon })\nonumber\\
&&\gamma_n(R)=\delta_n(R)-\beta_n(R)\nonumber\\
&&\hat h_n= {1\over 2R}\int_0^Rdr(h_{n\epsilon }(r)  +
h_{n\epsilon }(-r) )\nonumber\\
&&<h_{n  }>= h_{n\epsilon }-\hat h_{n  }
\label{7.2}
\end{eqnarray}
We use the method of induction to show the existence of a sequence of
functions for a small but nonzero interval of time, such that
\begin{equation}
||h_{n\epsilon }||_{H_1(-\infty , \infty )}\le {1\over (C^*-(4k'-1)\tilde Ct)
^{1\over 4k'-1}},
\label{7.3}
\end{equation}
where $\tilde C$ is the same constant that appears  in Eq. (\ref{7.7}),
$k'=\sup (1, k)$
and $(C^*)^{-1/(4k'-1)}=||h(t=0)||_{H_1(-\infty , \infty )}$.
Thus, $\tilde C$ and $C^*$ are some constants that depend only on initial
data, $k$ and coefficients of the polynomial $W$.

i) step $n=0$ is trivial. $h_0$ is at least $C^{1/2}$ as a function
of $R$ and $t$ and it obviously satisfies the bound.  Coefficients of
({7.3}) are $C^{3/2}$, thence there
exists a solution $h_{1\epsilon }\epsilon C^{1/2}$, by a standard result
for linear equations   \cite{Pietrovski} and   Proposition A.

ii) let there exists a solution $h_{n\epsilon }
\epsilon H_1(-\infty ,\infty )$
for some $n$. One easily infers that $h_{n\epsilon }$ satisfies the
conditions of the preceding Proposition, so that $h_n$
is $C^{1/2}$ as a function of $R$ and $t$.
Notice that $\delta_n(t, R) \le 1$. That means that the support of $h_n$
at a time $t$ must be placed within $-R^0-t, R_0+t$, that is, it remains
bounded.

 There exists also a short interval
of time such that $\delta_n(t, R)$ is positive, since initially
$\delta_n(0, R)=\delta (0, R) >0$.  We prove that using the induction
hypothesis. By direct computation one shows that
$$\partial_0\delta_n={-8\pi \over R}\int_0^Rdrr^2\Bigl[
W_n\partial_0\beta_n +\beta_nW'_n\partial_0\hat h_n\Bigr]
+{1\over R}\int_0^Rdr\partial_0\beta_n$$
and, from the definition of $\beta_n$ and $W_n$ and the approximating
equation (\ref{7.1a}),
$$\partial_0\beta_n =-16\pi \beta_n(\int_R^{\infty }+ \int_{-R}^{-\infty })
{dr\over r}<h_n>\Bigl[ A_n(r)-{1\over 2r} \int_{-r}^rd\tilde r A_n(\tilde r)
 \Bigr] $$
where
$$A_n =-\delta_{n-1}\partial_Rh_n +<h_n>\Bigl( 8\pi \beta_{n-1}RW_{n-1}
+{|R|^{\epsilon }\gamma_{n-1}\over R}\bigr) +\beta_{n-1}RW'_{n-1}/2.$$
One can bound  $\hat A_n$ by  $C R^{1/2}||h_{n\epsilon }||^x_{ H_1}$  and
$|\partial_0\beta_n(R)|$
and   $|\partial_0\delta_n| $ by $C ||h_{n\epsilon }||^x_{ H_1}$,
using the estimates  of   Lemmae 1 -5.
($C$ is a constant that depends only on initial data and may change from line
to line and $x$ is a number depending only on $W_n$ )
That shows, using the induction hypothesis on the    behaviour of Sobolev
norms of $h_{n\epsilon }$, that $\beta_n $  and $\delta_n$ are nonzero and
finite for a sufficiently small time $t$, if their initial values are nonzero.

Then various differentiability properties of
$<h_{n\epsilon }>, \hat h_{n\epsilon }, \beta_n, \gamma_n$ and $\delta_n$
follow immediately from  Lemmae 1-5 and  Steps 1-3 of  Lemma 1.
In particular,   the coefficient $\delta_n(t, R)$
is easily shown to be $C^{3/2}$ while the right hand side of the
approximating equation is certainly at least $C^0$; that guarantees the
existence of $h_{n+1\epsilon }$, thanks to a standard existence theorem
for linear equations as formulated by, for instance, Petrovsky. The
boundedness of $W'(\hat h_{n\epsilon })$ is
controlled due to estimates of $\hat h_{n\epsilon }$ and the assumption
that $W'(x)$ is bounded by a polynomial in $x$ with bounded coefficients.
We shall show that the $ H_1$ norm of  $h_{n+1,\epsilon }$ is bounded by
a number that depends only on initial data and $W$; that would
mean also that the interval of the existence of $h_{n\epsilon }$
is bounded from below
by a number that does not depend on the index $n$. In order to do so,
let differentiate the equation (\ref{ 7.1})   with respect $R$.
That gives an equation of the form
\begin{eqnarray}
&&(\partial_0 +\delta_{n} \partial_R)\partial_Rh_{n+1,\epsilon }
= {d\over dR}
\Bigl[ <h_{n\epsilon }>
(8\pi \beta_{n} RW_{n}+ {\gamma_{n}|R|^{\epsilon } \over R} )  +
{\beta_{n}R\over 2}W'_{n}\Bigr]-\nonumber\\
&&(\partial_Rh_{n+1\epsilon }){d\over dR}\delta_{n}.
\label{7.4}
\end{eqnarray}
Multiplying that equation by ${d\over dR}h_{n+1\epsilon }$, integrating over
the whole real
line and integrating by parts, one arrives  at
\begin{eqnarray}
&&\partial_0{1\over 2}||\partial_Rh_{n+1\epsilon }||^2
_{L_2(-\infty , \infty )}    =
\int_{-\infty }^{\infty } {d\over dR}h_{n+1,\epsilon } dR
\Bigl[ ({d\over dR} <h_{n\epsilon }>)(8\pi \beta_{n} RW_{n}+
{\gamma_{n} |R|^{\epsilon }\over R})
+ {d\over dR}({\beta_{n}R\over 2}W'_{n})+\nonumber\\
 &&  <h_{n\epsilon }>{d\over dR}(8\pi \beta_{n} RW_{n}+
  {\gamma_{n}|R|^{\epsilon } \over R} )
-\nonumber\\
&& {1\over 2}(\partial_Rh_{n+1\epsilon }){d\over dR}\delta_{n}\Bigr]
\label{7.5}
\end{eqnarray}
One can use the estimates of  Lemmae 1 - 5 and  Steps 1-3 and
 eventually arrive at the inequality
\begin{eqnarray}
&&{d\over dt}||h_{n+1\epsilon }||^2_{H_1(-\infty , \infty )} \le
\nonumber\\
&&C||h_{n+1,\epsilon }||_{H_1(-\infty , \infty )}\Bigl(
||h_{n}||^{4k}_{H_1(-\infty , \infty )} + ||h_{n\epsilon }
||^4_{H_1(-\infty , \infty )},
\Bigr)
\label{7.6}
\end{eqnarray}
where $C$ depends only on $k$ and initial data. Introducing a new constant
$\tilde C$ and $k'=\sup (4k, 4)$, one gets  the following inequality
\begin{eqnarray}
&&{d\over dt}||h_{n+1,\epsilon }||_{H_1(-\infty , \infty )} \le
\nonumber\\
&&\tilde C|
||h_{n\epsilon }||^{4k'}_{H_1(-\infty , \infty )}.
\label{7.7}
\end{eqnarray}
Using the induction
hypothesis and integrating (\ref{7.7}), one arrives at
\begin{equation}
||h_{n+1\epsilon }||_{H_1(-\infty , \infty )} \le
   {1\over (C^*  -(4k'-1)\tilde Ct)^{{1\over 4k'-1}}}
\label{7.8}
\end{equation}
which concludes the proof of the induction hypothesis. (\ref{7.8})
shows that the Sobolev norm of each function $h_{n\epsilon}$ is bounded by
$n$ - independent number and, that the interval $T$ of the existence of
solutions of all approximating equations is bounded away from zero by
a number that is $n$ - independent, $0<T<{C^*\over \tilde C(4k'-1)}$.
 From the approximating equation  and (\ref{7.8})
one deduces that $\int_{-\infty }^{\infty }
dr (\partial_0h_n(r, t))^2 \le C$, where $C$ is $t-$ and $n-$ independent.
Therefore $h_{n\epsilon }$ satisfies conditions of Proposition A, which
 implies   that the  sequence $h_{n\epsilon }$ is equicontinuous and
 equibounded.

Obviously, also
$\int_0^Tdt \int_{-\infty }^{\infty }
dr||h_{n\epsilon }||^2+ \int_0^Tdt\int_{-\infty }
^{\infty }dr|\partial_th_{n\epsilon }|^2\le C$ for some $C$.
 Now, the standard compactness argument shows the existence of
 a   subsequence $h_{n_i\epsilon }$ weakly convergent to    $h_{\epsilon }$
 in $ H_1([0, T]\times R)$.
$h_{n_i\epsilon }$ is equicontinuous and equibounded, therefore  by
the  Arzela-Ascoli theorem it contains a subsequence convergent
pointwise  to a limit  $h_{\epsilon }$. $h_{\epsilon }$ in turn, being a
limit of functions satisfying conditions of Proposition A, must be of class
$C^{1/2}$.
  The  pointwise convergence to $h_{\epsilon }$ and $C^{1/2}$
  continuity of $h_{\epsilon }$ implies the pointwise
 convergence of $\delta_n$, $\gamma_n$, $\beta_n$ and $\hat h_n$
  to functionals depending on the limiting solution $h_{\epsilon}$.
Thus the right hand side of (\ref{7.1a}) tends pointwise  to
an expression depending on the weak limit $h_{\epsilon}$.
%
%
We can conclude that $h_{n_i\epsilon }$ tends to a weak (distributional)
solution  $h_{\epsilon}$ of the equation (\ref{ 7.1}).

The norm of $h_{\epsilon}$ is bounded by a
constant  that depends on ${1\over \epsilon }$, so it would
become infinite when removing regularization, that is if
$\epsilon \rightarrow 0$. We will show, however, that there exists a
subset of initial data which gives rise to an evolution that
survives the removal of regularization.

\vskip 0.5cm

{\bf Part B.}

 Let initial data be  of   of compact support and
  $\partial_R h\epsilon L_4(-\infty , \infty )$; that implies
  also that $h\epsilon H_1(-\infty ,\infty )$.   Thus, by the result
  proven in Part A, there exists a local evolution. Now
one can show that $\partial_Rh_{\epsilon } \epsilon L_4(-\infty , \infty )$
 for  some time $0<t<T$.

One easily shows that in such a case  all estimates of  Lemmae 1-5
improve by a factor $R^{1/4}$. We have, in particular,
\begin{equation}
|\partial_R<h_{\epsilon}>| \le C{||\partial_rh_0||_{L_4(-\infty ,
\infty )} \over R^{1/4}}
\label{h_e}
\end{equation}
for any $t< T$ and with $C$ being (possibly) $\epsilon -$dependent.

In such a case we can improve, however, the statement  of Lemma 2, to get

 {\bf  Lemma 7.}
 Let   $\partial_r h_{\epsilon}\epsilon L_4(-\infty , \infty )$.
 Then  a solution of
 the regularized equation satisfies the following estimates
\begin{eqnarray}
&&|<h_{\epsilon}(t)>|\le
CR^{3/4}||\partial_rh_{\epsilon}(t)||_{L_4(-\infty , \infty )}\nonumber\\
&&||R^{\eta -1/2}\partial_R<h_{\epsilon}(t)>||_{L_2(-\infty , \infty }
\le C||h_{\epsilon}||_{H_1(-\infty , \infty )},
\label{h7}
\end{eqnarray}
where $C$ is $\epsilon -$independent.

With this new estimate one can show  that  $H_{1, 4}$ and $H_1$   norms of
$h_{\epsilon }$  remain uniformly bounded for $\epsilon \rightarrow 0$.
    Take a sequence of
$\epsilon_i$ tending to 0 as $i\rightarrow \infty $; there exists a
subsequence of  $h_{\epsilon_i}$ that is weakly convergent in $H_1$
to a limit  $h$; that is the sought solution of the equation (\ref{ 7.1}),
as can be shown by repeating arguments used in the final part of Part A.
  Also, $||\partial_Rh||_{L_4}<C$.   That accomplishes the
proof of Theorem 6.
\vskip 2cm

\centerline{V. Uniqueness of solutions.}

{\bf Theorem 8.} Under conditions of Theorem 6, if $W$ and $W'$
are Lipschitz continuous, there exists
a unique Cauchy solution of the reduced equation  (\ref{6.13}).

{\bf Proof.}

Let $h_1$ and $h_2$ be two solutions satisfying given initial data
of class $H_1\cap H_{1, 4}$.  We have $h_1(t=0, R)=
h_2(t=0, R)$.

 Let the suffix "1" or "2"  means that a function in question
$\beta , \gamma , \delta ,\hat h, <h>$ depends on
$h_1$ or $h_2$, respectively. Notice that $<f>+<g>=<f+g>$. We have
\begin{eqnarray}
\beta_1 (R, t) = e^{-8\pi (\int_R^{\infty }+\int_{-R}^{-\infty })
{<h_1>^2 \over r}}=\beta_2 e^{-8\pi (\int_R^{\infty }+\int_{-R}^{-\infty })
{<h_1+h_2> <-h_1+h_2>  \over r}}.
\label{beta1}
\end{eqnarray}
We can prove

{\bf Lemma 9}. Under conditions of Theorem 8,
\begin{equation}
|(\int_R^{\infty }+\int_{-R}^{-\infty }dr {<h_1+h_2> <-h_1+h_2>  \over r}|
\le C||h_1-h_2||_{L_4(-\infty ,\infty )}.
\label{f}
\end{equation}
Indeed, using several times the Schwarz inequality, the inequality
$(a-b)^2\le 2a^2+2b^2$
and the improved (for $h_i\epsilon H_{1,4}$) estimate of  Lemma  7
$$| <h>(R)| \le CR^{3/4} ||\partial_rh||_{L_4(-\infty ,\infty )}^2,$$
 one   bounds the integral of (\ref{f}) by
\begin{eqnarray}
&&\Bigl[ \int_{-\infty } ^{\infty }dr {<h_1+h_2>^2\over r^2}dr\Bigr]^{1/2}
\Bigl[ \int_{-\infty } ^{\infty }dr <h_1-h_2>^2 dr\Bigr]^{1/2} \le
\nonumber\\
&&C \Bigl[ \int_{-\infty } ^{\infty }dR \Bigl( (h_1-h_2)^2+{1\over 4R^2}
[\int_{- R }^R (h_1-h_2)dr]^2 \Bigr) \Bigr]^{1/2} \le \nonumber\\
&&C||h_1-h_2||_{L_4(-\infty , \infty )}.
\label{L4}
\end{eqnarray}
In the above calculation we used the finiteness of the support of initial
data; the constant $C$ depends on the support of initial data and on
$H_{1,4}$ norms of $h_1$ and $h_2$.

The above lemma yields, for small values of $||h_1-h_2||_{L_4(-\infty ,
\infty )}$ the following estimation
\begin{eqnarray}
|\beta_1 (R, t)-\beta_2 (R, t)|\le C||h_1-h_2||_{L_4(-\infty , \infty )}
 \label{beta2}
\end{eqnarray}
In a similar way one shows that
\begin{equation}
|(\hat h_1-\hat h_2)|\le {C\over R^{1/4}}
||h_1-h_2||_{L_4(-\infty , \infty )}
\end{equation}
and  (using Lipschitz continuity)
\begin{equation}
|W(\hat h_1)-W(\hat h_2)|\le {C\over R^{1/4}}
||h_1-h_2||_{L_4(-\infty , \infty )}.
\end{equation}
An analogous relation holds
for the difference $|W'(\hat h_1)-W'(\hat h_2|$.

>From (\ref{beta2}), (\ref{6.10} ) and the $W$ and $W'$
estimates, one shows that
\begin{eqnarray}
|\delta_1 (R, t)-\delta_2 (R, t)|\le C||h_1-h_2||_{L_4(-\infty , \infty )}.
 \label{delta1}
\end{eqnarray}
Similarly one arrives at
\begin{eqnarray}
|\gamma_1 (R, t)-\gamma_2 (R, t)|\le C||h_1-h_2||_{L_4(-\infty , \infty )}.
 \label{gamma1}
\end{eqnarray}

 Above and below $C$ is a certain constant that changes from  line to line,
independent of $t$ and $R$.

Substracting the reduced equations for $h_2$ from that for $h_1$
and using the above estimates on the right hand side of the substracted
equations, one gets  (below $\Delta h= h_1-h_2$)
\begin{equation}
(\partial_0 +\delta_1 \partial_R  )\Delta  h  +(\delta_1-\delta_2)
\partial_Rh_2 \le C|\Delta h|+ \Bigl( |<h_1>|+|h_2|\Bigr)
||\Delta h||_{L_4(-\infty , \infty )});
\label{final1}
\end{equation}
 multiplying (\ref{final1}) by $(\Delta h)^3$, once again estimating the
 difference $|\delta_1-\delta_2|$ by  $||\Delta h||_{L_4(-\infty , \infty )}$
 and integrating by parts, one eventually arrives at the inequality
\begin{equation}
\partial_0||\Delta h||_{L_4(-\infty , \infty )}^4\le
C||\Delta h||_{L_4(-\infty , \infty )}^4;
\label{final2}
\end{equation}
that implies $||\Delta h||_{L_4(-\infty , \infty )}=0$, since at $t=0$
$\Delta h=0$.
The last inequality holds true for sufficiently small $t$.
$h_1$ and $h_2$ are continuous functions, therefore
$h_1=h_2$ at least for sufficiently small intervals of time. Iteration
of that reasoning leads to the conclusion that
if there exists a solution
of the reduced equation,
then it is unique in the $L_{\infty }$ norm.

That means, in turn,
that the possible nonuniqueness can be seen on the level of first
derivatives of   $h$ and, even if there exist two solutions with different
derivatives, then still   $\gamma_1=\gamma_2$,
$ \hat h_1=\hat h_2 $, $ \delta_1 =\delta_2$ and $\gamma_1=\gamma_2$ up to
their first derivatives.

Using that  one can easily show that also the $H_1$ norm of the difference
$ \Delta h$ must vanish. In fact, let $dh = \partial_R\Delta h $;
from the reduced equation one gets
\begin{equation}
\partial_0dh= - \partial_R(\delta dh) +dh F
\label{unique7}
\end{equation}
where $F $    denotes terms which do not involve $dh$.
Integrating (\ref{unique7}) over $R$ one gets, after employing various
estimates proven in the first part of this paper
\begin{equation}
\partial_0||dh||_{L_2(-\infty , \infty )}\le
C||dh||_{L_2(-\infty , \infty )};
\label{unique8})
\end{equation}
that yields $||dh||_{L_2(-\infty , \infty )}=0$, from Gr\"onwall
inequality, since at $t=0$ we have  $||dh||_{L_2(-\infty , \infty )}=0$.
Combining that with the already proven fact,
we conclude that  the solution of the reduced equation is unique in the
sense of $H_1$.  A similar reasoning gives uniqueness in $H_{1,4}$.
 That ends the proof of Theorem 8.
\vskip 2cm

\centerline{VI. External Cauchy problem.}

It occurs that that there are two main problems
in proving the global  existence .

One is due to difficulties in estimating
needed quantities at the origin. That we omit by considering a sort
of an external Cauchy evolution;
 we will investigate whether   initial data  of Einstein - scalar field
equations give rise to an evolution that exists globally outside any
outgoing null hypersurface   originating at  an  initial
spacelike hypersurface.

The other is the possible emergence of minimal
surfaces during an evolution; that would mean that equations become singular.
We will show that this does not happen; that fact is  well known
and it shows  that polar gauges  are deficient in the sense that they
do avoid regions of spacetime with minimal surfaces; if
initially minimal surfaces are absent, then
they cannot develop in Cauchy slices satisfying the condition
$trK=K_r^r$ during a finite evolution, assuming that a dominant energy
condition is satisfied. The proof of this claim goes as follows. Let
 $\Omega^{out}_{R'}=[(R, t): |R|\ge R_{in},
{dR_{in}\over dt}=\delta , R_{in}(t=0)=R'>0]$
  be a patch of hypersurfaces
that evolve from an initial slice $\Sigma_{ R'}^{out}$ (which is free of
minimal surfaces) and let $\Sigma_{R_0t}^{out}$ be the first slice with
a minimal surface located at an areal radius $R_m$.
$R_{in}(t) $ describes the location of the free inner boundary. By the
regularity of the evolution, the four-metric is at least $ C^1$
on that piece of the space-time; thus there exists a null ingoing
geodesic joining the four-point $(R_m,t)$ with a point $(R'>R_m, 0)$
lying on the initial slice. Along that geodesic the mean curvature
$p$ decreases from an initial nonzero value $p_0$ to 0. The  change
of the mean curvature along the ingoing null geodesic  is given, however,
by one of the Raychaudhuri equations (that can be obtained, in that case,
by manipulating the evolution Einstein equation and the hamiltonian
constraint). We use the geodesic coordinates  with the line element
$ds^2 =-N^2dt^2+ dl^2 +R^2d\Omega^2$
One can find, after some calculations, that
\begin{equation}
(\partial_t-  N \partial_l)(pR)= 8\pi NR (j+\rho )+{N\over 4R}(p^2R^2-4);
\label{VI.1}
\end{equation}
using now the energy condition $|j|\le \rho $,
definition of the lapse $N$
and the fact that $R$ is lowering along the
ingoing null ray, one gets the inequality
\begin{equation}
(\partial_t-  N\partial_lR)(pR)\ge  - {pR\beta (R)\over 2 R_m}\ge
- {pR \over 2R_m}.
\label{VI.2}
\end{equation}
In the last line I used the estimation $\beta (R)\le 1$. Equation
(\ref{VI.2}) yields
\begin{equation}
 pR\ge \inf_{\Sigma_{ R'0}^{out}} (pR)e^{-t/2R_m},
\label{VI.3}
\end{equation}
which must be nonzero for $t<\infty $. Thus we obtained a contradiction, that
enforces us to accept that   polar gauge  slicings cannot penetrate
regions with minimal surfaces.   Notice also that (\ref{VI.3}) gives a
lower bound  for the minimal value of mean curvature on subsequent
Cauchy slices; that will used later.

Now we  state the main  result of this section.

{\bf Theorem 10.}
 Take a part
$\Sigma_{R'}=[(R, t=0):|R|\ge R', ] $
(with $R'\ge 0$) of the initial hypersurface $\Sigma_0$ .
  Let   initial data of  equation (\ref{6.13})
 on an initial slice  be  of compact support, $\inf_{\Sigma_{R'}} \delta >0$,
 the mass function $m(R') \le m$ (with $h(R')=h(-R')=0$ if  $m(R')= m$)
 and

i) $  h_0\epsilon H_{1,4}(\Sigma_{R'} )$; assume also that

ii) $(\int_{-\infty }^{-R'}+ \int_{R'}^{\infty }) dr h_0(R, t)=0$.
 Let $0\le W(x)$  and $|W'(x)|$ be bounded by a polynomial with constant
 coefficients of order $k $. Then there exists a global Cauchy  solution
in $\Omega^{out}_{R'}$.

Theorem 10  has been proven in \cite{Malec1995}, under stronger
differentiability  conditions, $h\epsilon H_2$. Below we present a modified
proof that bases on the results of preceding sections.

First of all, we have to  write down the reduced  problem in  a
 modified (but equivalent) form.  While keeping $\beta  $ in the
 form (\ref{6.12}),  we choose  the  representation (\ref{6.9}) of $\delta $,
 namely
\begin{equation}
\delta (R)=   {(pR)^2\over 4}\beta (R)=
\Bigl( {1-{2m\over |R|}+{2m(R)\over |R|}}\Bigr)\beta (R)  ,
\label{VI.4}
\end{equation}
where $m$ is the asymptotic mass and
\begin{equation}
m(R)=4\pi \int_R^{\infty }drr^2\rho = 4\pi \Bigl( \int_R^{\infty }dr
-\int_{-R}^{-\infty }dr \Bigr)\Bigl( {\delta (r)\over \beta (r)}  <h>^2
+{r^2\over 2} W(\hat h)\Bigr) ,
\label{VI.5}
\end{equation}
It is convenient to deal with $\hat h$ expressed as follows
\begin{equation}
\hat h(R) ={-1\over 2R}
\Bigl( \int_{-\infty }^{-R}+ \int_R^{\infty } \Bigr) h(r)dr
\label{VI.6}
\end{equation}
which is equivalent to the expression  (\ref{6.12}) used before, if
$\int_{-\infty }^{\infty }dr h(r) =0$.
With those forms of $\beta ,\delta , h$ and $\gamma $  it is obvious that
solutions of the reduced equation (\ref{6.13}) outside any given
outgoing null cone $\delta_{R'}$ do not depend on  its interior.
We use  that fact in proving Theorem 10.

Namely, we smoothly extend initial data across $R'>0$ to vacuum,
keeping  conditions $\int_{-\infty }^{\infty }h(r)dr =0 $, $m= m(0)$
and $\delta >0$,
 to get $h(r)=0$ for $|r|<R'-\eta $ for some $\eta >0$;
 it is easy to  see that there exist extensions
which do not change significantly the required $H_1$ and $H_{1,4}$
Sobolev norms. Therefore we may use the local result of Theorem 6 to infer
the existence of a local solution; notice that according to the
preceding remark,   outside
an outgoing null cone $\delta H_{R'}$  (including the cone itself)
defined by ${dR\over dt}=
\delta , R(t=0)=R'$, the solution  is independent of the extension.

There is a number of useful local estimates; obviously $0<\beta \le 1$,
$0< \delta\le 1 $, $\gamma \le 2$  and $m(R)\le m$. We need a bound
on $\hat h$.  That is proven in the following

{\bf Lemma 11.} Under conditions of Theorem 6,
\begin{equation}
|\hat h(R)|\le
{\sqrt{m}\over \sqrt{4\pi R} (\inf_{R'\ge R}{\delta \over \beta})^{1/2} }
\label{VI.7}
\end{equation}
Proof of Lemma 11. (Assume, for simplicity, $R>0$).

We have  $\hat h(R) =-\int_R^{\infty }dr \partial_r\hat hdr=
(\int_R^{\infty }+
\int^{-R}_{-\infty }){<h>\over r}dr$;  using the Schwartz  inequality, we
get
$$|\hat h(R)| \le \Bigl[ (\int_R^{\infty }+
\int^{-R}_{-\infty })dr<h>^2
\int_R^{\infty }dr{1\over r^2}\Bigr] ^{1/2}$$
which is bounded by
$${1\over R^{1/2}(\inf_{R'\ge R}{\delta \over \beta})^{1/2}}
\Bigl[  (\int_{R}^{ \infty }dr +
\int^{-R}_{-\infty }){\delta (r)\over
\beta (r)}\Bigl( <h>^2+{1\over 2} W(\hat h)\Bigr) \Bigr]^{1/2}.$$
The integral term is bounded from above by $ { m(R) \over 4\pi }$
which in turn is not bigger than ${ m \over 4\pi } $.
That ends the proof of Lemma 11.

Take now   the patch of slices of constant $t$ of $\Omega^{out}_{R'}$.
    It is easy to find, manipulating with the reduced
equation  (\ref{6.13}) that the rate of change of the
$L_p$ norm of $h$ (for  any  even value of $p$) along the external
foliation is bounded,
\begin{equation}
{d\over dt}||h||_{L_p(\Sigma_{R't}^{out})}\le C
||h||_{L_p(\Sigma_{R't}^{out})},
\label{VI.8}
\end{equation}
 where $C$ depends on the above local estimates. $C$ can be infinite if
 minimal surfaces appear, but  that cannot happen for $t<\infty $,
 as  proven at the beginning of this  section. Therefore the growth of
 $L_p$ norm is controlled. A similar reasoning gives a control also
 of $H_1$ and $H_{1,4}$ norms.

 The bootstrap argument yields now immediately the global existence.
 Indeed, let $T$ be the maximal existence interval of a solution in the
 exterior region: thus at any $T-\eta $ all norms are finite. By using
 the above reasoning one shows that relevant norms must be finite at $t=T$,
 which leads to contradiction. That ends the proof of Theorem 10.

 {\bf Remark on smoothness.}
 In the external region one can reduce the smoothness
 requirements on $h$ from $H_{1,4}$ to $H_1$. Indeed, if initial data
 vanish on a compact neigbourhood of the symmetry
 center on the initial hypersurface, then there exists a compact
 domain $[-R(T), R(T)]\times [0, T]$ with null data. In such a case the
 regularization procedure of Section IV is not necessary and one can show the
 existence of local solutions in $H_1$. In the globalization part presented
 above  one uses only those local extensions that are null close to the
 origin $R=0$; therefore also the global existence extends to $H_1$.
 That reasoning allows one to conclude that Theorem 10 holds true also
 for matter with selfinteraction $W$ that is singular at the origin but
 satisfies the  remaining boundedness conditions. Thus there exists a global
 evolution in an external region
 for Yang Mills $SU(2)$ fields (with $W={(1-\phi ^2)^2\over 2R^2}$)
and skyrmionic  $SU(2)$ fields (with $W= { \sin^2(\phi ) \over 2R^2}$).

{\bf Remarks on the free boundary $\delta H_{R'}$.}
 It is easy to notice (see (\ref{VI.5})
 that outside the Schwarzschild region,
 $R'\ge 2m$, minimal surfaces must be absent
 ($\delta $ or $pR$ must be strictly positive). In that case the
 inner boundary $\delta H_{R'}$ of $\Omega^{out}_{R'}$
 escapes to spatial infinity, $|R_{in}(t)|$ increases without bound.
 Therefore initial data  posed  outside the Schwarzschild radius
 always give rise to global external  solutions.

 In the alternative case, with the initial hypersurface entering
 the interior of the Schwarzschild sphere, we may consider
 two situations:

i) a rather trivial  case, when the inner boundary  $\delta_{R'}$ of
some of the future Cauchy slices crosses
(at some finite $t'$) through the sphere located
at the areal radius $2m$; in that case we have the global existence, with the
conclusion, that $\delta H_{R'}$ escapes to spatial infinity.

ii) the inner boundary "freezes" close to a sphere of an areal radius
$R< 2m$.

We will  investigate the second point in more detail.
 One can easily show that the area of an outermost
apparent horizon cannot decrease (see, e. g. \cite{MOM1994a}); in fact it has
to increase whenever matter (satisfying the strong energy condition)
crosses through the horizon;  that has to
move acausally outwards. Asymptotically the areal radius
of the apparent horizon becomes equal to $2m_B$, where
$m_B$ is the Bondi mass  of the black hole.
Take a part $\Sigma_{r0}^{out}$  of the initial hypersurface  that
does not include minimal surfaces. Then  data on  $\Sigma_{r0}^{out}$
give rise to
a local evolution, according to the local    Theorem 6.   The global
evolution prolongs  until the free  inner boundary  freezes at
some areal radius $R<2m$, close to the (anticipated) minimal surface.
In  such a case one can take a slightly smaller initial open end
$\Sigma_{r'}^{out}\subset
\Sigma_r^{out}$; that evolves to a spacetime that  freezes at a later time
than the previous one. Continuing that procedure ad infinitum  one finds
finally a smallest open end such that
   the area $dH$ of  a null inner boundary $\delta_{R'}$  a corresponding
   spacetime $H$  still    stabilizes at a value $4\pi R_B^2$.
$\delta_{R'}$ is an event horizon and half of $R_B$ is the Bondi mass.
 Thus   there    exists a solution
for that exterior region  $\Sigma_{R't}^{out}$ whose inner boundary coincides
with an  event horizon that is asymptotic to a minimal surface
located somewhere at $R\le 2m$.  That solution is global in the sense that
it does exist for  arbitrarily large $t$, but on the other hand it does not
cover a
part of the physical spacetime which is hidden behind an apparent horizon.
\vskip 2cm

\centerline{VII. Global existence and central integral singularities.}

{\bf Theorem 12.} Assume conditions of Theorem 6.  Assume that there
exists a small cylinder $\Omega^{R_0} =[(R, t): R\le R_0]$  such that
a     contribution to $H_1\cap H_{1,4}$ norm of $h$ from a spatial section
$t=const$, $\Omega^{R_0}_t$, of $\Omega^{R_0} $ is uniformly bounded,
$||h||_{ H_{1,4}(\Omega^{R_0}_t)}<C$,
where $C$ is $t$ - independent.
Then the Cauchy evolution of the Einstein-scalar field system exists
globally.

 {\bf Proof.} In the first part of the proof we will use the global existence
 of solutions of the  related    Stefan problem.
Using Theorem 10, the proof of Theorem 12 proceeds as follows.
Take $\Omega^{out}_{R_0/2}$; by the proof of Theorem 10, the norm of
$h$ in $\Omega^{out}_{R_0/2}$ is bounded by at most exponentially increasing
function of time,  $||h||_{H_1(\Omega^{out}_{  R_0/2,t})}<C_0e^{ct}$.
Thus, taking into account the uniform bound in $\Omega^{R_0}_t$,
  we have    $||h||_{H_1(-\infty , \infty )}<\infty $ at a time $t= R_0/2$.
Now, take a portion $\Sigma_{t, R_0/2}$ of the Cauchy slice at a time t;
using the same reasoning as before, we can extend the  existence period from
$R_0/2$ into $R_0$. Iterating that reasoning we infer the global existence.

If we assume a condition stronger than in Theorem 12, namely
that (keeping the same notation)
$$\sup_{0<R_0<2m}\Bigl[
{1\over R_0^{1/2}}||h||^2_{  H_{1,4}(\Omega^{R_0}_t)}
\Bigr] < C,$$
where $C$ is small enough  then the spacetime is geodesically complete.
For definiteness, consider the  massless scalar fields;
then $C={1\over 48 \pi }$. Indeed, from (\ref{VI.4})
and  (\ref{VI.5})  one obtains
\begin{eqnarray}
&&\delta (R) = \beta (R)\Bigl( 1 -
{8\pi \over R} \int_0^Rdr
 \Bigl( {\delta (r)\over \beta (r)}  (<h(R)>^2+
 <h(-R)>^2  )\Bigr) \ge  \nonumber\\
&&\beta (R)\Bigl( 1-{1\over 48 R^{1/2}\pi }||h||^2_{L_4(\Omega^{R_0}_t)}
\Bigr) \ge
\nonumber\\
&&\beta (R)\bigl( 1-{1\over 48 R^{1/2}\pi }||h||^2_{H_{1,4}(\Omega^{R_0}_t)}
\Bigr) >0.
\label{VII.1}
\end{eqnarray}
The first inequality follows from $<h>^2\le 2h^2+2\hat h^2$,
$\int_R^Rdr h^2dr\le 2R^{1/2}||h||_{L_4(-R, R)}$ and
$\int_R^Rdr\hat h^2dr\le R^{1/2}||h||_{L_4(-R, R)}$.  (\ref{VII.1}) means
that all time-like and null-like geodesics have infinite proper or
affine length.

{\bf Remark.} Theorem 12 essentially states that if there is no central
singularity (understood as a portion of spacetime that gives infinite
contribution to the $H_{1, 4}$ norm of $h$), then there is no singularity
at all. That is an accordance with a corresponding result proven by
Rein, Rendall and Schaeffer \cite{rendall95} in the case of the
Vlasov - Einstein system.

That conclusion is consistent with
an analogous result in \cite{MOM1994}, proven
entirely in the framework of initial data formalism,
which shows the absence of   $L_2$    singularities
 on any Cauchy slice with a regular trace of extrinsic curvature and
with (at most ) a conical singularity in the symmetry center.

\vskip 2cm

\centerline{VIII. Naked singularities.}

We will give an example of an initial configuration that
gives rise to $H_{1,4}$ evolution and  that is characterized by
a pointwise curvature singularity at the symmetry center. That singularity
can be seen
by  external observers  placed at spatial infinity.
 Other examples of naked singularities in various material systems
can be found in \cite{Ioshi}.

As a material model we choose  a scalar field with the
nonlinear selfinteraction potential
\begin{equation}
W(\phi ) = \sin^2(\phi ).
\label{VIII.1}
\end{equation}
Assume that $h(R)= 0$ outside some $|R_0|$,
$h(R)=\lambda \epsilon (R) |R|^{\alpha }$ for $|R|<R_0$ and ${3\over 4}
<\alpha <1$ (where $\epsilon (x) =1 $ for $x>0$ and $-1$ for $x<0$),
with a smooth transition
in between. Then $\hat h$ vanishes identically in the initial hypersurface,
$\beta (R, t=0) =e^{-{8\pi \lambda^2\over \alpha }\Bigl( (R_0)^{2\alpha }
-R^{2\alpha }\Bigr) }\ge e^{-{8\pi \lambda^2\over \alpha }(R_0)^{2\alpha }}$
and $\delta (R) = {1\over R}\int_0^R\beta (r) dr\ge
e^{-{8\pi \lambda^2\over \alpha }  (R_0)^{2\alpha }}$.
 The energy density $\rho $ at $t=0$ is
equal to $4\pi {\delta \over R^2\beta }h^2$ and it is divergent like
$R^{-2+2\alpha }$ at the origin. The hamiltonian constraint (\ref{2.2})
yields now that the three-dimensional Ricci scalar $^{(3)}R$ is also
divergent like $R^{-2+2\alpha }$.

 The $H_{1,4}$ norm of  $h$ is finite and it is merely
proportional to  ${\lambda \over 4\alpha -3}$.  Therefore there exists
an evolution in an interval $T$. From the local analysis of Section IV
one obtains
\begin{equation}
||h  ||_{H_{1,4}(t)} \le
   {1\over \Bigl(  C^*  -(4k'-1) \lambda  Ct\Bigr)^{{1\over 4k'-1}}}
\label{VIII.2}
\end{equation}
where $k'\ge 1, (C^*)^{1\over 4k'-1}=
{1\over ||h  ||_{H_{1,4}(t=0)}} $ and  $C$ is a certain
constant. Therefore the smaller is $\lambda $, the bigger is  the existence
interval $T$.

By differentiation of the metric function $\delta $ one obtains
 $$\partial_0\delta ={-8\pi \over R}\int_0^Rdrr^2\Bigl[
W \partial_0\beta  +\beta W' \partial_0\hat h \Bigr]
+{1\over R}\int_0^Rdr\partial_0\beta $$
and, from the definition of $\beta $ and $W $ and the reduced
equation (\ref{6.13}),
$$\partial_0\beta  =-16\pi \beta (\int_R^{\infty }+ \int_{-R}^{-\infty })
{dr\over r}<h >\Bigl[ A (r)-{1\over 2r} \int_{-r}^rd\tilde r A (\tilde r)
 \Bigr] $$
where
$$A  =-\delta \partial_Rh  +<h >\Bigl( 8\pi \beta RW
+{ \gamma \over R}\bigr) +\beta RW' /2.$$
One can bound  $|\partial_0\beta (R)|$ by $C R||h  ||^{k'}_{ H_{1,4}}$
and then also $|\partial_0\delta | $ by $C ||h ||^{k'}_{ H_1}$,
using  estimates analogous to those
of  Lemmae 1 -5.  Combining that with (\ref{VIII.2})
yields
\begin{equation}
\kappa \equiv \inf_{R}\delta (R, t) \ge
e^{-{8\pi \lambda^2\over \alpha }  (R_0)^{2\alpha }} -C(\lambda )T,
\label{VIII.4}
\end{equation}
where $C(\lambda )\rightarrow 0$ for $\lambda \rightarrow 0$ and $t\le T$.

   Let  $\lambda $  be  so  small  as to have ${2m\over \kappa} <T$.
Then (since $\delta $ is at least $C^1$)
there exists a solution $R(t)$ of the null geodesic equation
\begin{equation}
{dR\over dt }=\delta
\label{VIII.5}
\end{equation}
with the initial value $R(0)=0$ such that $R(T)>2m$. The exterior of the
cylinder $R=2m$ is geodesically complete, therefore the  null geodesic
$R(t)$ reaches any symptotic observer. Hence we established that
the central curvature singularity is naked.

A closer investigation shall reveal that this singularity is not strong
in the sense of Tipler (\cite{Tipler}).

  That example  suggests   that  the
concept of {\it pointwise  singularities} shall be replaced
  by a  class of {\it quasilocal (integral) singularities}. The latter
  are understood as those local singularities that are characterized by
infinite values of some quasilocal (integral) quantities (e. g.,
some Sobolev  norms).  That notion  of a singularity seems to be  more natural
than the idea of strong singularities, since  quasilocal quantities
can be directly related to a quantitative description of the Cauchy evolution.
Consequently, the concept of the cosmic censorship shall be accordingly
reformulated.

 \vskip 2cm

Acknowledgements.

This work is supported by the KBN grant 2PO3B 090 08.

The author is grateful to Konstanty Holly and Jan Stochel from the
Institute of Mathematics of Jagellonian University and
Alan Rendall from AEI in Potsdam for discussions, advice
and useful comments.


\begin{references}
\bibitem{Choquet - Bruhat}  Y. Choquet - Bruhat, p. 130 in Gravitation:
an introduction to current research, L. Witten (editor), J. Wiley and Sons,
New York London 1962;
\bibitem{Chr-Kl} D. Christodoulou, S. Klainermann, The global nonlinear
stability of the  Minkowski space,  Princeton University Press,
Princeton 1993;
\bibitem{ (Christodoulou) } D. Christodoulou
{\it Commun. Math. Phys.} {\bf 105},
337(1986); {\bf 106}, 587(1987); {\bf 109}, 613(1987);
{\it Ann. of Math.} {\bf 140}, 607(1994);
\bibitem{rendall95} G. Rein, A. D. Rendall and J. Schaeffer,
{\it Commun. Math. Phys.} {\bf 168}, 467(1995);
\bibitem{Penrose} R. Penrose,  {\it Riv. N. Cimento}, {\bf 1}, 252(1969);
p. 631 - 668 in Seminar on Differential Geometry, Princeton: Princeton
University Press 1982;
\bibitem{Moncrief} V. Moncrief and D. Eardley,
{\it Gen. Rel. Gravitat.} {\bf 13}, 887(1981);
\bibitem{Malec1996}  E. Malec, in preparation;
\bibitem{Wald} R. Wald, General Relativity, Chicago University Press 1984;
\bibitem{Godunov}  S. K. Godunov, p. 122 -123 in  Uravneniya
matematiceskoj fiziki, Nauka, Moscow 1979 (in Russian).
\bibitem{Pietrovski} I. G.  Petrovsky, Lectures on Partial Differential
Equations, Warsaw 1955, PWN ( that is in Polish; there are also
numerous Russian and English editions);
\bibitem{Malec1995} E. Malec  {\it Classical and Quantum Gravity} {\bf 13},
1849(1996);
\bibitem{MOM1994a}   E. Malec and N. \'O Murchadha, {\it Phys. Rev.}
{\bf D49}, 6931(1994);
\bibitem{MOM1994}   E. Malec and N. \'O Murchadha, {\it Phys. Rev.}
{\bf D50}, R6033(1994);
\bibitem{Ioshi} D. Christodoulou, {\it Commun. Math. Phys.} {\bf 93},
171(1984); R. P. A. C. Newman, {\it Class. Quantum Grav.} {\bf 3},
527(1986); B. Waugh and K. Lake, {\it Phys. Rev.} {\bf D38}, 1315(1988);
A. Ori and T. Piran, {\it Phys. Rev.} {\bf D42}, 1068(1990);
I. H. Dvivedi and P. S. Joshi,  {\it Class. Quantum Grav.} {\bf 9},
L69(1992); P. S. Joshi and I. H. Dvivedi, {\it Commun. Math. Phys.}
{\bf 146}, 333(1992);   K. S. Virbhadra, S. Jhingan and P. S. Joshi,
{\it  Nature of singularity in Einstein-massless scalar theory},
gr-qc preprint 9512030(1995);
\bibitem{Tipler} F. J. Tipler, C. J. S. Clarke and G. F. R. Ellis,
p. 97 in   General  Relativity  and Gravitation, A. Held (editor),
Plenum, New York 1980.
\end{references}
\end{document}